\documentclass[twocolumn,aps,superscriptaddress,showpacs,nofootinbib,floatfix]{revtex4}
\usepackage{epsfig,bm,feynmf}
\usepackage{graphics}
\usepackage{amsmath}
\usepackage[normalem]{ulem}  
\usepackage[dvips]{color} 

\renewcommand\sout{\bgroup \color{red} \ULdepth=-.5ex \ULset}
\newcommand{\be}{\begin{equation}}
\newcommand{\ee}{\end{equation}}
\newcommand{\bea}{\begin{eqnarray}}
\newcommand{\eea}{\end{eqnarray}}
\begin{document}


\title{The production of primordial $J/\psi$ in p+p and relativistic heavy-ion collisions}


\author{Taesoo Song}\email{Taesoo.Song@theo.physik.uni-giessen.de}
\affiliation{Institute for Theoretical Physics, Johann Wolfgang Goethe Universit\"{a}t, Frankfurt am Main, Germany}
\affiliation{Frankfurt Institute for Advanced Studies, Johann Wolfgang Goethe Universit\"{a}t, Frankfurt am Main, Germany}
\affiliation{Institut f\"{u}r Theoretische Physik, Universit\"{a}t
Gie\ss en, Germany}

\author{Joerg Aichelin}
\affiliation{Subatech, UMR 6457, IN2P3/CNRS, Universit\'{e} de Nantes, \'{E}cole des Mines de Nantes, 4 rue Alfred Kastler, 44307 Nantes Cedex 3, France}
\affiliation{Frankfurt Institute for Advanced Studies, Johann Wolfgang Goethe Universit\"{a}t, Frankfurt am Main, Germany}

\author{Elena Bratkovskaya}
\affiliation{Institute for Theoretical Physics, Johann Wolfgang Goethe Universit\"{a}t, Frankfurt am Main, Germany}
\affiliation{GSI Helmholtzzentrum f\"{u}r Schwerionenforschung GmbH Planckstrasse 1, 64291 Darmstadt, Germany}

\begin{abstract}
$J/\psi$ observables of  p+p collisions at $\sqrt{s}=$ 200 GeV and 2.76 TeV are calculated using the PYTHIA event generator for the production of $c$ and $\bar c$ and the Wigner density formalism. The charm quark momentum from PYTHIA is tuned in such a way that the transverse momentum and rapidity distribution correspond to the Fixed-Order Next-to-Leading Logarithm (FONLL) calculations and hence to the experimental data for open charm mesons. Using the Wigner density of the charmonium wave function we  calculate for each charm quark pair the  probability to form a charmonium.  As a result, the experimental data on the total $J/\psi$ yield,  its rapidity and transverse momentum distribution as well as the fractions of feed-down from excited states of charmonium are reproduced. Applying the same approach to relativistic heavy-ion collisions, we find, if quark gluon plasma (QGP) effects are ignored, a considerable enhancement of the primordial $J/\psi$'s. 
\end{abstract}




\maketitle

\section{introduction}

Relativistic heavy-ion collisions have the purpose to produce extremely hot and dense matter which is supposed to be similar to the early universe and to the inner structure of neutron stars. This matter, once produced, forms a QGP which expands with a high velocity before it hadronizes. Therefore, the life time of a QGP is very short and to study its properties is a big challenge, theoretically as well as experimentally.

The study of the modifications of the spectra of J/psi is one of the promising probes to investigate the properties of the QGP since
Matsui and Satz have proposed that the suppression of $J/\psi$ is a signature of quark-gluon plasma formation~\cite{Matsui:1986dk,Zhao:2007hh,Linnyk:2008hp,Liu:2009gx,Song:2011xi,Zhou:2014kka}.

In order to extract  precisely the effect of the existence of a QGP on the quarkonium production in relativistic heavy-ion collisions, it is necessary to understand its production in elementary reactions like in p+p collisions. The quarkonium is not directly produced, but has its origin in the production of heavy quark pairs. This process requires a large energy-momentum transfer and can therefore be described in perturbative quantum chromodynamics (pQCD). The formation of a quarkonium from the heavy quark pair is, however, a soft process which cannot be described by pQCD. Presently only phenomenological models have been advanced.

The color singlet model (CSM) assumes that a quarkonium is formed only from the heavy quark pair which is produced in a color-singlet state and has the same angular-momentum quantum numbers as the quarkonium~\cite{Chang:1979nn,Berger:1980ni,Baier:1983va}. The CDF (Collider Detector at Fermilab) Collaboration has found, however,  that the CSM underestimates charmonium production in $p+\bar{p}$ collisions at $\sqrt{s}=$ 1.8 TeV by more than one order of magnitude~\cite{Abe:1997jz,Abe:1997yz}.

Non-relativistic quantum chromodynamics (NRQCD) is based on an effective Lagrangian keeping all necessary symmetries~\cite{Caswell:1985ui}. The Lagrangian is organized as a power series in inverse heavy quark mass.
The coefficient of each operator term is matched with the original QCD Lagrangian through Green functions or experimental or lattice data~\cite{Pineda:2011dg}. It can describe quarkonium decay and production by using factorization, and reproduces the experimental data well~\cite{Bodwin:1994jh,Shao:2014yta}.
Since both color singlet and color octet states of the initial heavy quark pair contribute to quarkonium production, it is sometimes called the color octet model.

In the color evaporation model (CEM), the heavy quark pair whose invariant mass is smaller than twice the heavy meson mass turns into a quarkonium. If the invariant mass of the pair is larger than twice the heavy meson mass, two open heavy-flavor hadrons are produced~\cite{Fritzsch:1977ay,Halzen:1977rs,Gluck:1977zm,Barger:1979js,Amundson:1996qr}.
Whether a quark pair is in a color-singlet state or in a color-octet state is not considered in CEM, because the color charge is assumed to evaporate during the process of quarkonium formation.
The fraction of the population of the different states of the quarkonium is a parameter which does not depend on collision energy.

Another approach to study the formation of quarkonia is the coalescence model. It was originally invented to explain the unexpected large number of deuterons observed in heavy ion collisions~\cite{Butler:1961pr,Butler:1963pp,Schwarzschild:1963zz} by the assumption that a neutron and a proton with a small relative momentum can form a deuteron under the emission of a photon. Whereas in Ref.~\cite{Butler:1961pr,Butler:1963pp,Schwarzschild:1963zz} the explicit matrix elements have been calculated, later coalescence parameter have been introduced \cite{Gutbrod:1988gt} which are adjusted to reproduce the deuteron data. Still later it has been realized that in the Wigner density formulation of quantum mechanics the coalescence in phase space is similar to the projection of the 2-body Wigner density onto the Wigner density of the bound state~\cite{Baltz:1995tv,Greco:2003xt,Greco:2003mm,Song:2016lfv}.
This approach has been applied in heavy ion collisions to deuteron formation ~ \cite{Remler:1981du,Gyulassy:1982pe,Aichelin:1987rh} as well as to the formation of bound states of quarks and antiquarks~\cite{Baltz:1995tv,Greco:2003xt,Greco:2003mm,Zhang:2006yf, Kahana:2010md,Han:2016uhh}. It allowed to formulate the formation of bound states as a dynamical process. Most of these studies are focused on the regeneration of $J/\psi$ in relativistic heavy-ion collisions. Since the wave function of typical hadrons peaks at small relative momenta, quarks and antiquarks with a small invariant mass have a higher probability to form a bound state.  In this sense, the Wigner density approach for quarkonium production is similar to CEM.

However, the Wigner density approach considers not only the relative momentum but also the spatial distance between coalescence partners. The latter may not be so important in elementary scattering.
In relativistic heavy-ion collisions, however, many particles are produced very closely.
The production of charm quarks is not an exception. Such a compact production may affect the formation of primordial quarkonium in relativistic heavy-ion collisions because a charm and an anticharm quark from different primary vertices may form a charmonium.

Whether in the QGP, created in heavy ion collisions, stable charmonia states exist is still an open question. It has been found that the yield of $J/\psi$ can also be obtained assuming a statistical hadronisation \cite{BraunMunzinger:2000px, BraunMunzinger:2000ep,BraunMunzinger:2009ih}. 

This paper is organized as follows:
In Sec.~\ref{coalescence} we briefly recall the Wigner density approach.
It is applied to quarkonium production in p+p collisions, and the results are compared with experimental data in Sec.~\ref{pp}.
The same approach is then applied in Au+Au collisions at 200 GeV and in Pb+Pb collisions at 2.76 TeV in Sec.~\ref{hic}, and the summary follows in Sec.~\ref{summary}.

\section{Our model}\label{coalescence}
Our model to describe the initial distribution of quarkonia is based on the assumptions that the production of color singlet quarkonia is a very fast process because the many-body interaction among the partons created in p+p collisions acts only for a very short time. This time is much shorter than the time of a collision with asymptotic free states. Under this condition we can apply the sudden approximation which states that the probability amplitude $a_i$  that an initial $c \bar c$ pair, described by the relative wave function $|\Psi>$, forms a quarkonia of type i with the relative wave function $|\Phi_i>$ is given by
\be
a_i= <\Psi|\Phi_i>\nonumber\\
\ee
where  $|\Phi_i>$ is the eigenstate i of the interacting color neutral $c \bar c$ system in free space. Since the initial relative wave function
$|\Psi>$ of the $c\bar c$ pair is not known, we construct it here from the output of the PYTHIA event generator, which gives only the momenta of the particles, assuming that they are classical.

Since the potential depends on the relative distance of the $c$ and  $\bar c$ only and treating the heavy quarks nonrelativistically, the pair wave function $| \phi_i>$ is a product of a plane wave of the center of mass motion and the wave function of the relative coordinates $| \Phi_i>$. The density matrix of the quarks in the two-body state
\be
|\Phi_i><\Phi_i|
\ee
can be converted into a Wigner density
\be
\Phi^W_i({\bf r},{\bf p})=\int d^3y e^{i{\bf p\cdot y}}<{\bf r}-\frac{1}{2}{\bf y}|\Phi_i><\Phi_i|{\bf r}+\frac{1}{2}{\bf y}>.
\ee
The Wigner density has the normalization $\sum_i \int d^3rd^3p\ \Phi^W_i({\bf r},{\bf p})=1$.
\be
n_i({\bf R,P}) =\int  d^3r d^3p\  \Phi^W_i({\bf r},{\bf p})  n^{(2)}({\bf r_1,p_1,r_2,p_2})
\ee
is the probability density to find a $c \bar c$ pair in the eigenstate $i$, if $n^{(2)}({\bf r_1,p_1,r_2,p_2})$ is  the two body density matrix of the system in Wigner representation. Here we have used
\begin{eqnarray}
{\bf R}=\frac{{\bf r_1}+{\bf r_2}}{2},~~~{\bf r}={\bf r_1}-{\bf r_2},\nonumber\\
{\bf P}={\bf p_1}+{\bf p_2},~~~{\bf p}=\frac{{\bf p_1}-{\bf p_2}}{2}.~~
\label{define1}
\end{eqnarray}
If the probability density is small, this approach can be extended to N- body systems
\bea
n_i({\bf R,P}) &=&\sum \int \frac{ d^3r d^3p}{(2\pi)^3}\  \Phi^W_i({\bf r},{\bf p}) \prod_j \int \frac{ d^3r_j d^3p_j}{(2\pi)^3} \nonumber \\
&& n^{(N)}({\bf r_1,p_1,r_2,p_2,...,r_N,p_N})
\eea
where the sum runs over all possible $c \bar c $ pairs in the N - body Wigner density with relative coordinates {\bf r}, {\bf p} and center of mass coordinates {\bf R}, {\bf P}. The product stands for all the other than the pair coordinates. The total probability that a $c \bar c$ pair is formed is given by
\be
P_i=\int \frac{ d^3R d^3P}{(2\pi)^3} n_i({\bf R,P})
\ee
whereas
\be
\frac{dP_i}{d^3P}=\int \frac{ d^3R}{(2\pi)^3} n_i({\bf R,P})
\ee
gives the spectrum of the  $c \bar c $ pairs.
Using the wavefunction of a simple harmonic oscillator (SHO),
the Wigner density of a  $S-$state and a $P-$state are, respectively, given by~\cite{Baltz:1995tv,Cho:2011ew},
\begin{eqnarray}
\Phi^W_{\rm S}({\bf r, p})&=&8\frac{D}{d_1 d_2}\exp\bigg[-\frac{r^2}{\sigma^2}-\sigma^2p^2\bigg],\label{wigner1}\\
\Phi^W_{\rm P}({\bf r, p})&=&\frac{16}{3}\frac{D}{d_1 d_2}\bigg(\frac{r^2}{\sigma^2}-\frac{3}{2}+\sigma^2p^2\bigg)\nonumber\\
&&\times\exp\bigg[-\frac{r^2}{\sigma^2}-\sigma^2p^2\bigg],
\label{wigner2}
\end{eqnarray}
where $\sigma^2=2/3\langle r^2\rangle$ for $S-$state and $\sigma^2=2/5\langle r^2\rangle$ for $P-$state with $\sqrt{\langle r^2\rangle}$ being the root-mean-square (rms) radius, and
$D$, $d_1$, and $d_2$ are the color-spin degeneracies of meson, quark and antiquark, respectively.
We note that $r$ is not the radius but the diameter of quarkonium based on the definition in Eq.~(\ref{define1}).
\begin{figure}[!h]
\centerline{
\includegraphics[width=9.5 cm]{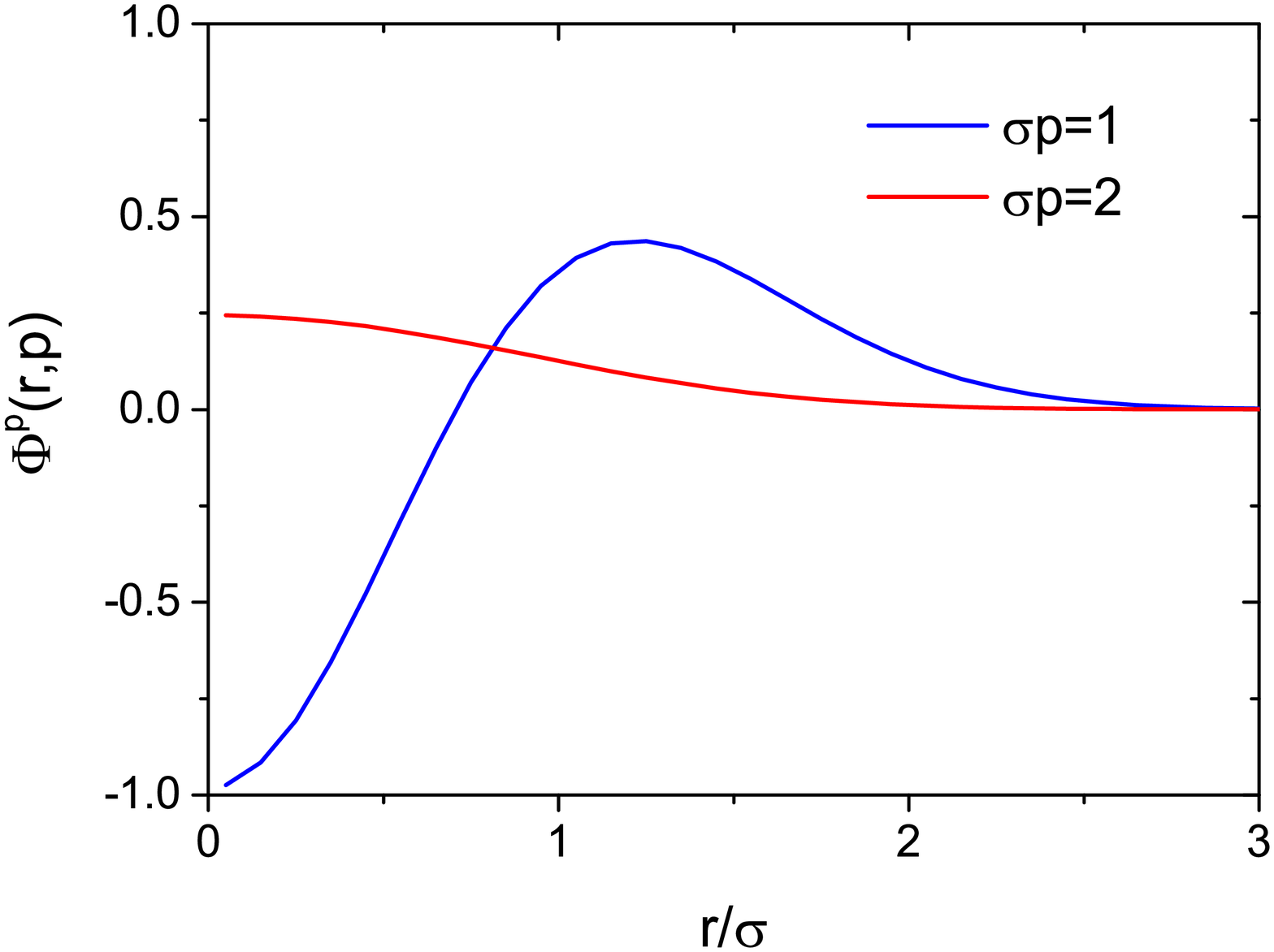}}
\caption{(Color online) the Wigner function for a $P-$state as a function of $r/\sigma$ for $\sigma p=$1 and 2.}
\label{wigner-p}
\end{figure}

The Wigner density for a $S-$state has the maximum value at $r=0$ and $p=0$.
However, the Wigner density for a $P-$state is negative at the point due to two reasons:
Firstly,  because both position and momentum cannot simultaneously be specified in quantum mechanics and secondly
because two particles located at the same point cannot have nonzero angular momentum in the center-of-mass frame.
Figure \ref{wigner-p} shows the Wigner function for a $P-$state as a function of $r/\sigma$ for $\sigma p=$1 and 2.
As mentioned above, the Wigner function is negative at small $r$ for small $p$.
The maximum probability is found at
\begin{eqnarray}
r^2=\sigma^2\bigg(\frac{5}{2}-\sigma^2p^2\bigg).
\label{pmax}
\end{eqnarray}
However, if $\sigma^2p^2$ is larger than $5/2$,
the Wigner function is always positive at physical $r$ ($r\ge 0$).
In this study, if the Wigner density is negative for a ceratin charm quark pair, it is turned to zero.

The initial relative wave function or Wigner density of the $c \bar c$ pair, $|\Psi>$, cannot be calculated from first principles. We assume that it can be approximated by a Gaussian Wigner density around the initial relative distance r and the relative momentum p of the pair particles:
\be
W_\Psi(\mathbf{r,p})= C e^{-r^2\mu^2}e^{-p^2 /\mu^2}
\label{wavefct}
\ee
where $\mu $ is proportional to the reduced mass of the $c \bar c$ pair: $\mu \propto m_c/2$ with $m_c$ being charm quark mass. The distribution of the relative momentum of the $c \bar c$ pair in PYTHIA is then the convolution of the initial relative momentum distribution with Eq. (\ref{wavefct}). The single particle spectrum  of $c$ or $\bar c$ quarks is therefore that of PYTHIA.

Since the ensemble of the eigenstates $\{\Phi_i\}$ of the $c \bar c$ pair is a basis of the two particle Hilbert space, the expectation value of the energy of the initial state can be calculated and is given by:
\bea
<\Psi|H|\Psi> &=& \sum_i<\Psi|\Phi_i><\Phi_i|H|\Phi_i><\Phi_i|\Psi>\nonumber \\
& =& \sum_i  E_i  a_i^2.
\eea

\section{$J/\psi$ production in p+p collisions}\label{pp}

In order to study quarkonium production in p+p collisions, we first need the initial distribution of the charm quark pairs.
The theoretical state-of-the-art transverse momentum and rapidity distributions of charm quarks are those calculated in fixed-order next-to-leading logarithm (FONLL) approximation~\cite{Cacciari:1998it,Cacciari:2001td} which have been confirmed by experiments.
In our approach the charm quarks are produced by the PYTHIA event generator. Though  PYTHIA takes into account the initial and final state showers, we have to tune the charm quark momentum and rapidity from the PYTHIA such that transverse momentum and rapidity distribution are similar to those from the FONLL.

\begin{figure}[!h]
\centerline{
\includegraphics[width=9.5 cm]{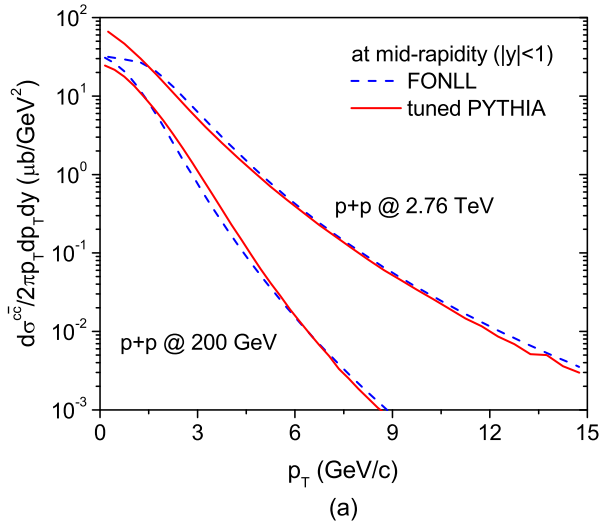}}
\centerline{
\includegraphics[width=9.5 cm]{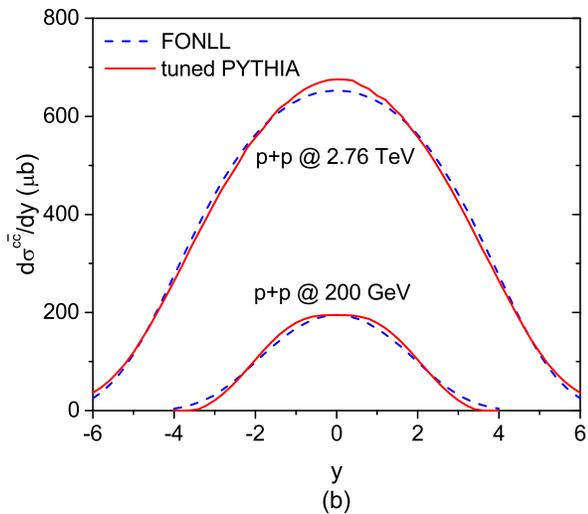}}
\caption{(Color online) the transverse momentum at $|y|<1$ and rapidity distribution of charm quarks in p+p collisions at $\sqrt{s}=$ 200 GeV and 2.76 TeV from the FONLL and from the tuned PYTHIA calculations.}
\label{ccbar}
\end{figure}

Figure~\ref{ccbar} compares the transverse momentum and rapidity distributions of charm quarks in p+p collisions at $\sqrt{s}=$ 200 GeV and 2.76 TeV from the FONLL calculation with those from the PYTHIA calculations after tuning. This requires
that at $\sqrt{s}=$ 200 GeV the transverse momentum and the rapidity of charm quarks are both reduced by 15 \%.
At $\sqrt{s}=$ 2.76 TeV only the rapidity has to be reduced by 9 \% from the Innsbruck tune~\cite{Song:2015sfa,Song:2015ykw}. We can see that with these corrections both, transverse momentum and rapidity distributions, are in good agreement with FONLL.

\begin{figure}[!h]
\centerline{
\includegraphics[width=9.5 cm]{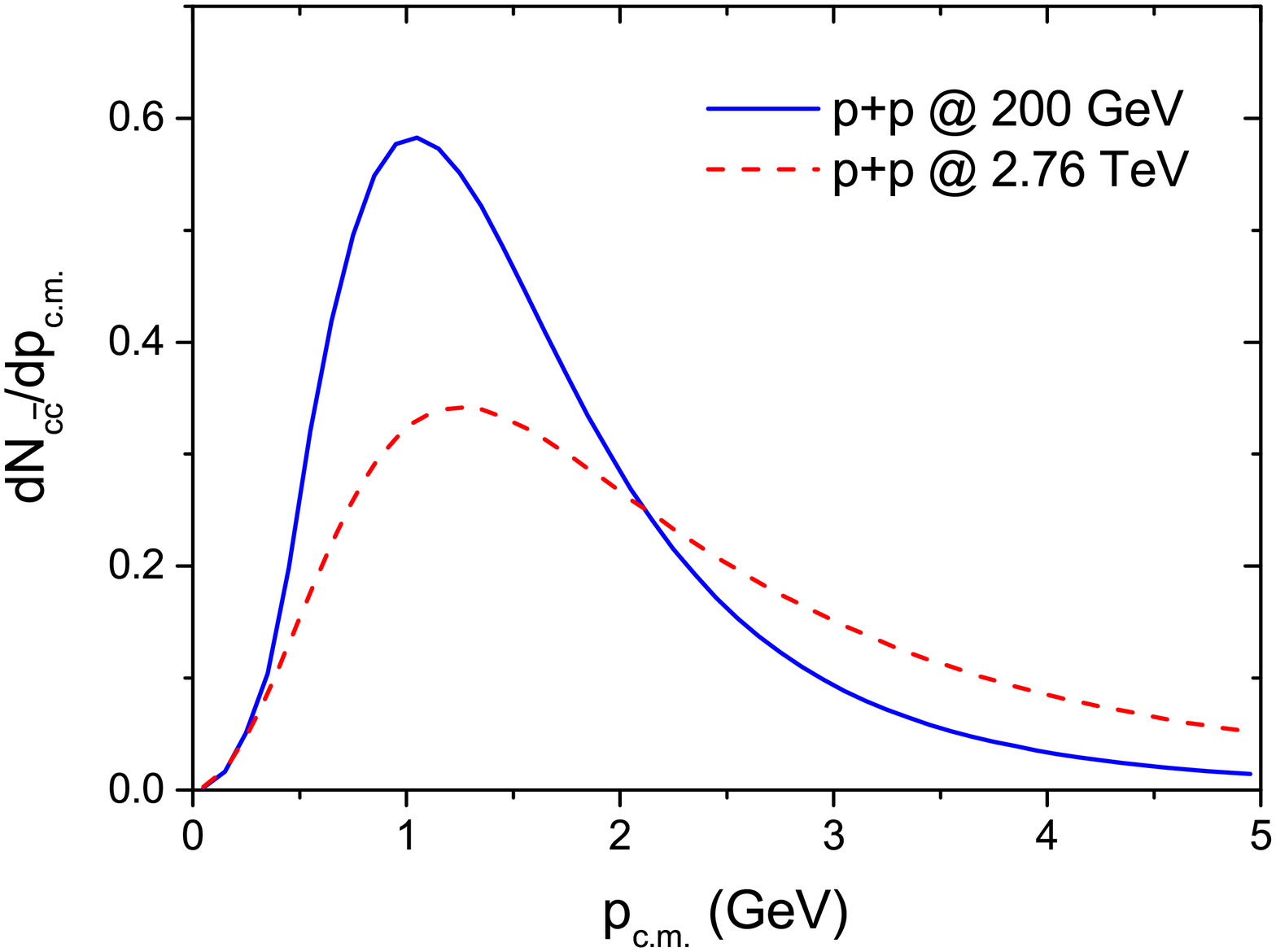}}
\caption{(Color online) The distribution  of the relative momentum in the center-of-mass frame of charm quark pair from the tuned PYTHIA calculations for p+p collisions at $\sqrt{s}=$ 200 GeV and 2.76 TeV.}
\label{dist}
\end{figure}

Now we come to the production of charmonium from the charm quark pairs.
In Fig.~\ref{dist} we show the distribution of the relative momentum in the center-of-mass  of charm quark pairs
from the tuned PYTHIA calculations for  p+p collisions at $\sqrt{s}=$ 200 GeV and 2.76 TeV.
We can see that charm pairs have a larger relative momentum at $\sqrt{s}=$ 2.76 TeV as compared to at $\sqrt{s}=$ 200 GeV. CEM predicts therefore a lower charmonium production at the LHC as compared to RHIC. This is, however,  partially
compensated by the increasing contribution from bottom decay to $J/\psi$ ~\cite{CMS:2011ora}.

Considering the LO calculations in pQCD,
charm and anticharm quarks are produced close by in $q+\bar{q}$ annihilation and in the $s-$channel of $g+g$ scattering.
In the $t-$ or $u-$ channel of $g+g$ scattering, the production points are separated by a large virtuality.
We assume that the initial separation of charm and anticharm quarks in space has a gaussian distribution with the mean-radius-square of charm quark pair being the inverse of charm quark mass:
\be
W_\Psi(\mathbf{r,p})
\sim r^2 \exp\bigg(-\frac{r^2}{2\delta^2}\bigg),
\ee
where $\delta^2=\langle r^2\rangle/3=4/(3m_c^2)$ such that $\sqrt{\langle r^2\rangle}/2=1/m_c$.

\section{Results}

We come now to the results of our calculations.
\begin{figure}[!h]
\centerline{
\includegraphics[width=9.5 cm]{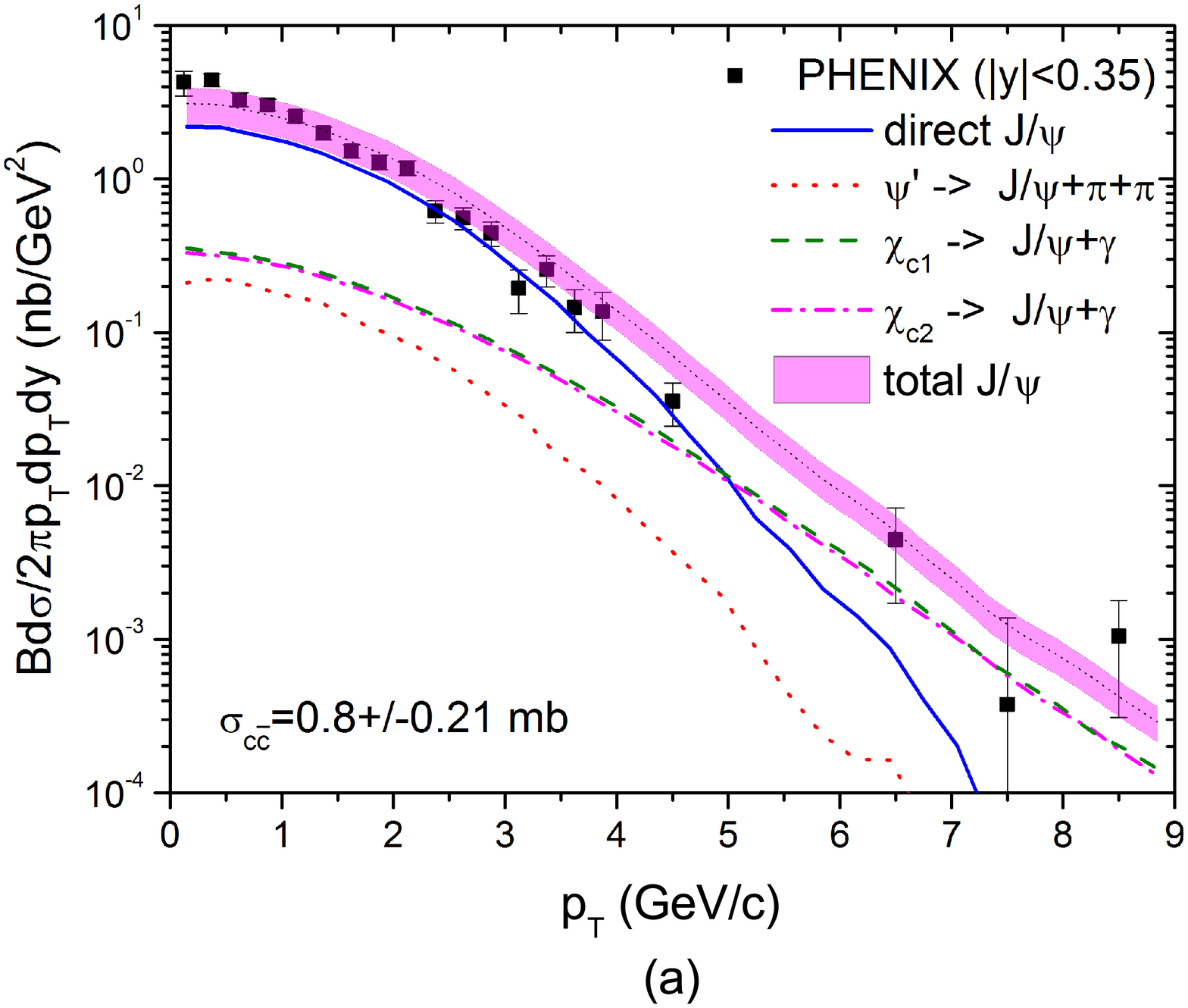}}
\centerline{
\includegraphics[width=9.5 cm]{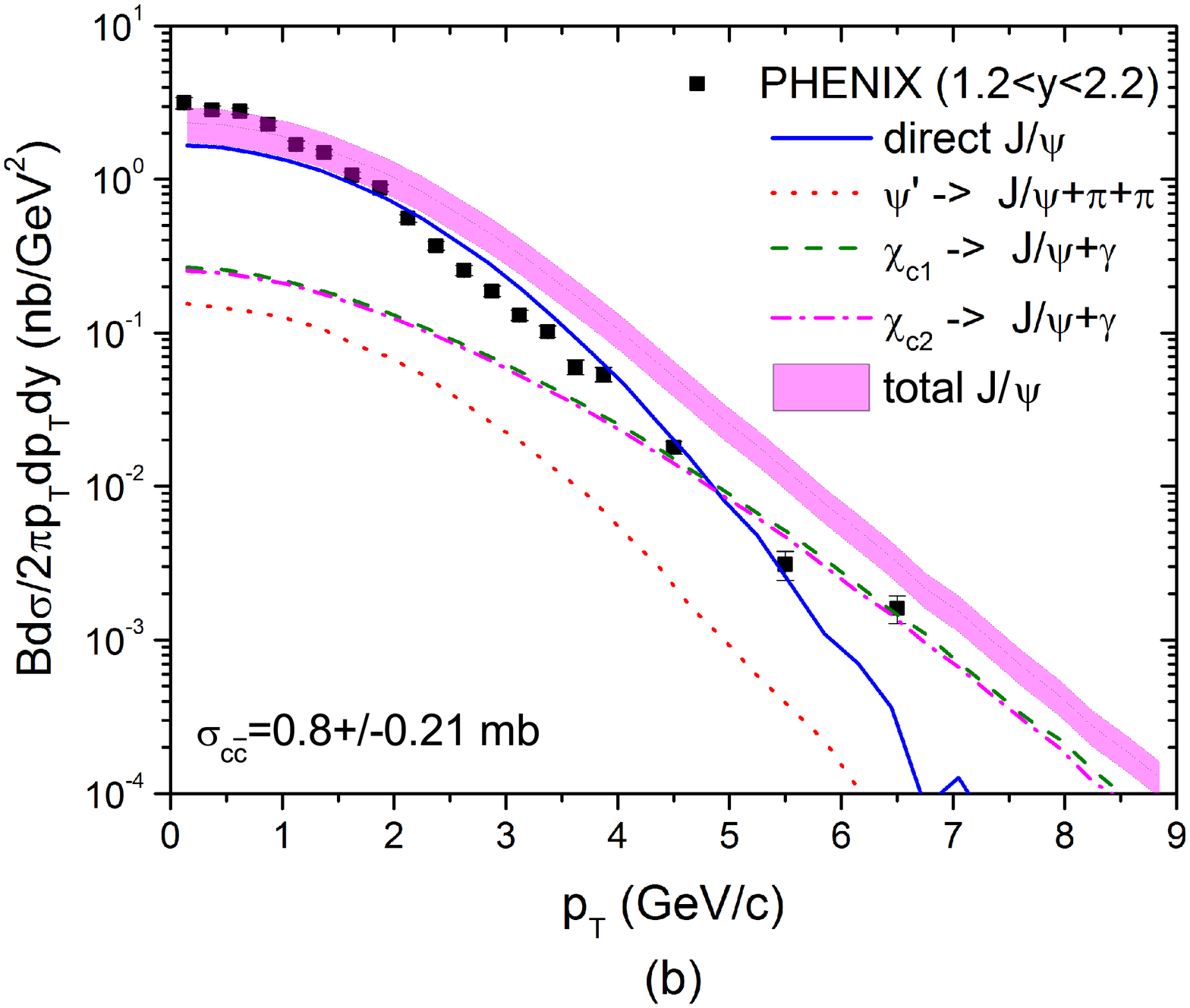}}
\centerline{
\includegraphics[width=9.5 cm]{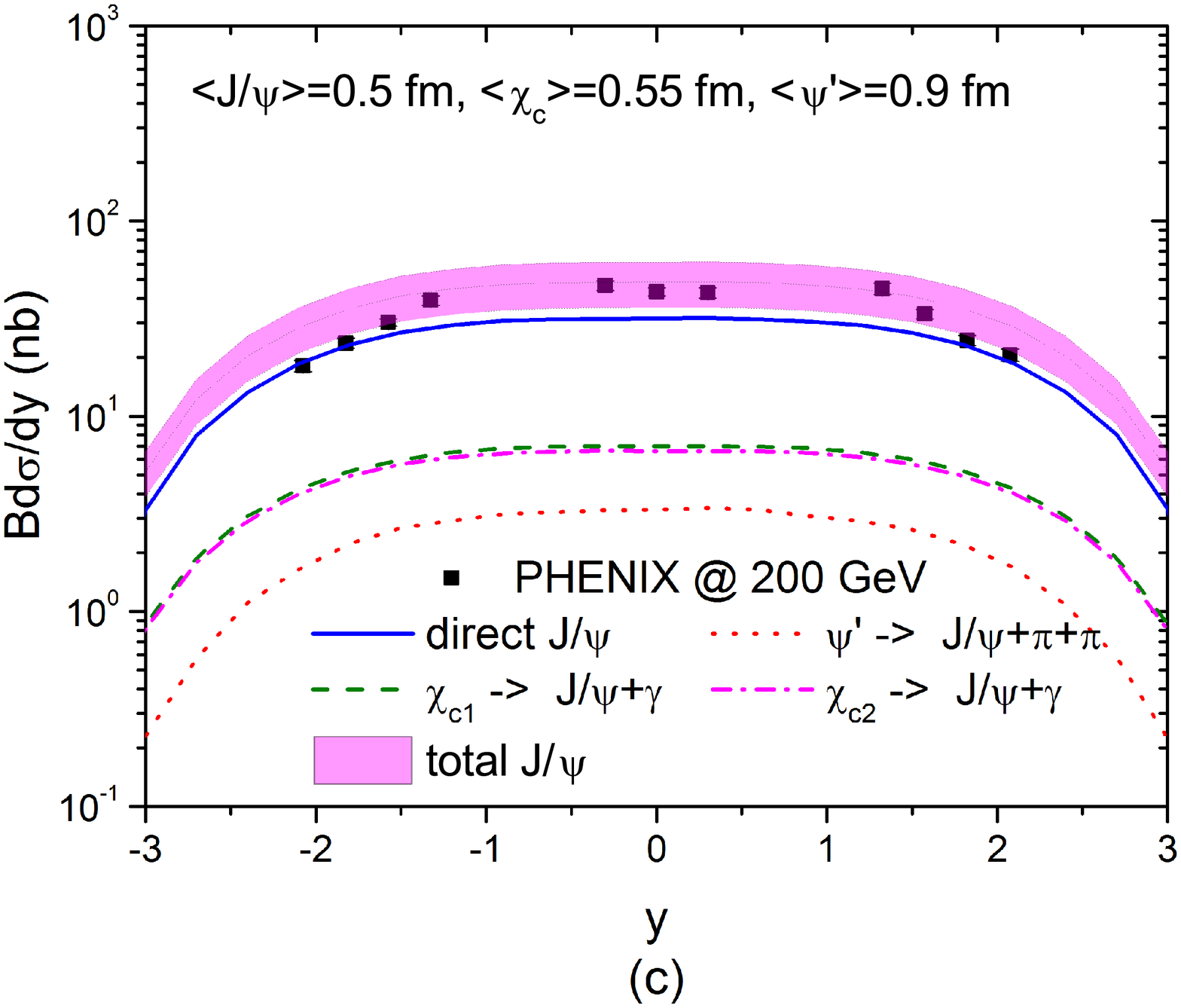}}
\caption{(Color online) $J/\psi$ production in p+p collisions at $\sqrt{s}=$ 200 GeV including the feed-down from $\chi_c$ and $\psi^\prime$ in comparison with the experimental data from the PHENIX Collaboration~\cite{Adare:2006kf}.}
\label{rhic}
\end{figure}

\begin{figure}[!h]
\centerline{
\includegraphics[width=9.5 cm]{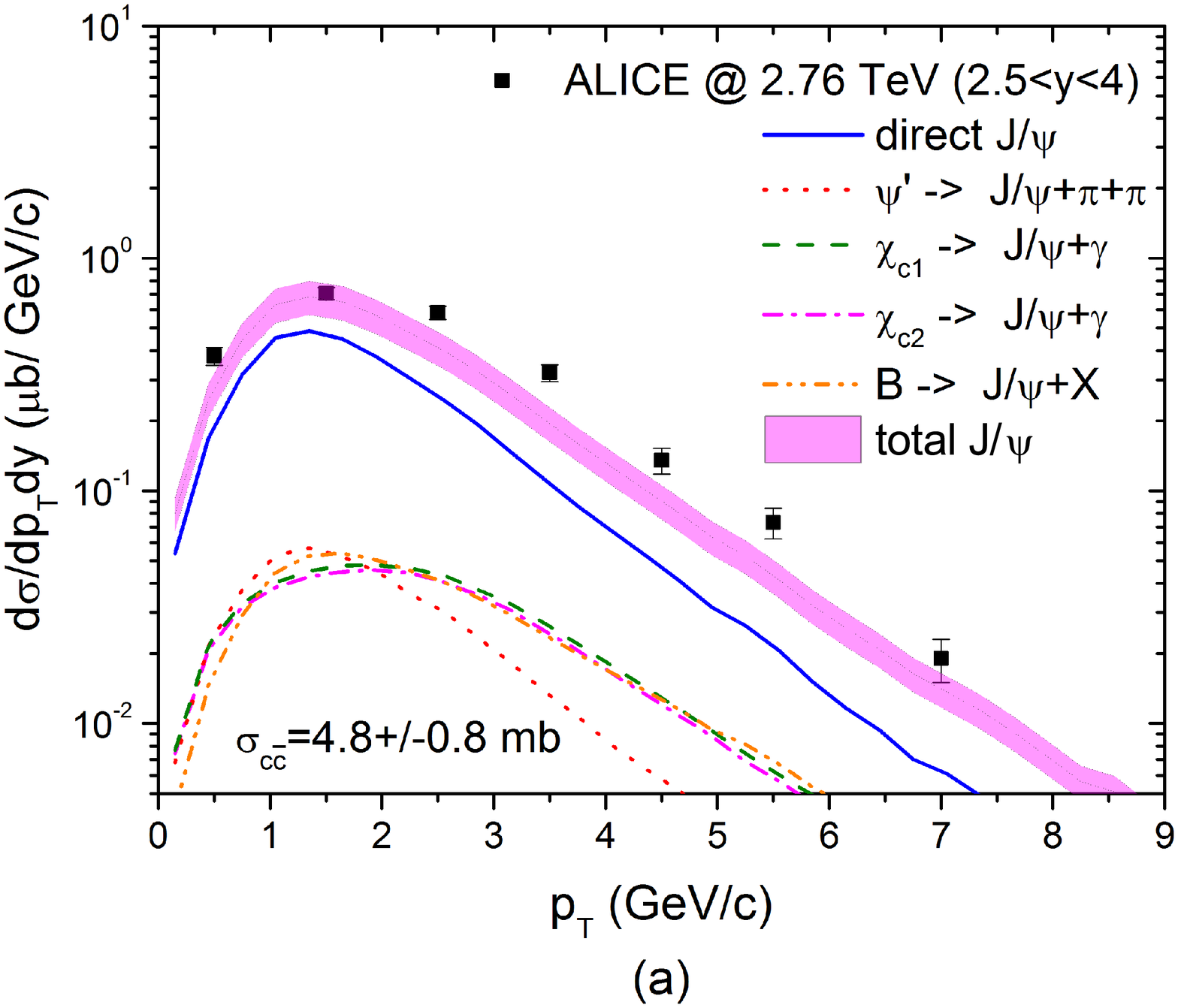}}
\centerline{
\includegraphics[width=9.5 cm]{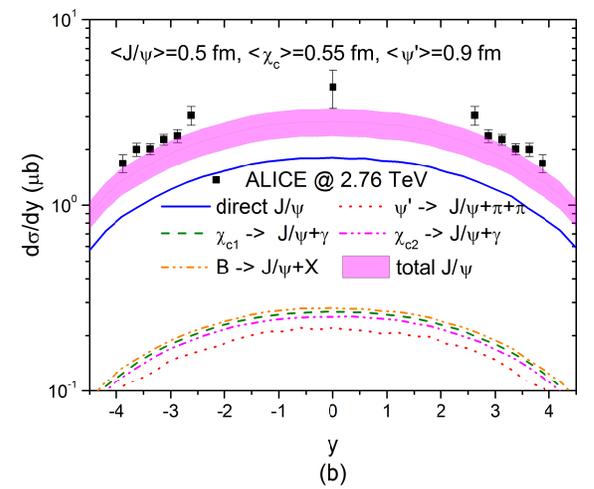}}
\caption{(Color online) $J/\psi$ production in p+p collisions at $\sqrt{s}=$ 2.76 TeV including the feed-down from $\chi_c$, $\psi^\prime$, and $B$ mesons in comparison with the experimental data from the ALICE Collaboration~\cite{Abelev:2012kr}.}
\label{lhc}
\end{figure}

Figure~\ref{rhic} and \ref{lhc} show the direct $J/\psi$ production in p+p collisions as well as that via  feed-down from  $\chi_c$ and $\psi^\prime$ at $\sqrt{s}$= 200 GeV and 2.76 TeV, respectively.
At  $\sqrt{s}=$ 2.76 TeV, the $J/\psi$ has an additional contribution from $B$ meson decay, which is parameterized as $0.05+0.02 p_T$ with $p_T$ being the transverse momentum of the $J/\psi$ in GeV~\cite{CMS:2011ora}.
The bands represent the total calculated $J/\psi$ yield and take into account the error bars of the experimental data on the scattering cross sections for charm production in p+p collisions.
The scattering cross sections are taken to be $0.8\pm0.21$ mb at $\sqrt{s}=$ 200 GeV~\cite{Adamczyk:2012af} and  $4.8\pm0.8$ mb at $\sqrt{s}=$ 2.76 TeV~\cite{Abelev:2012vra}.
$J/\psi$'s from $\chi_c\rightarrow J/\psi+\gamma$ have a harder $\rm p_T-$spectrum than that from $\psi^\prime\rightarrow J/\psi+\pi+\pi$, though $\psi^\prime$ is heavier than $\chi_c$. The reason is that the daughter particle $\gamma$ is much lighter than the two pions.
We don't show here the feed-down from $\chi_{c0}\rightarrow J/\psi+\gamma$, for its contribution is very small.
We see that our results are compatible with the experimental data from the PHENIX Collaboration~\cite{Adare:2006kf} and from the ALICE Collaboration~\cite{Abelev:2012kr}, for the rapidity distribution as well as for the $p_T$ distribution.
We overestimate slightly the $J/\psi$ production at forward and backward rapidities at $\sqrt{s}=$ 200 GeV. Maybe a better rapidity distribution of the initial charm quark pairs improves this.

\begin{figure}[!h]
\centerline{
\includegraphics[width=9.5 cm]{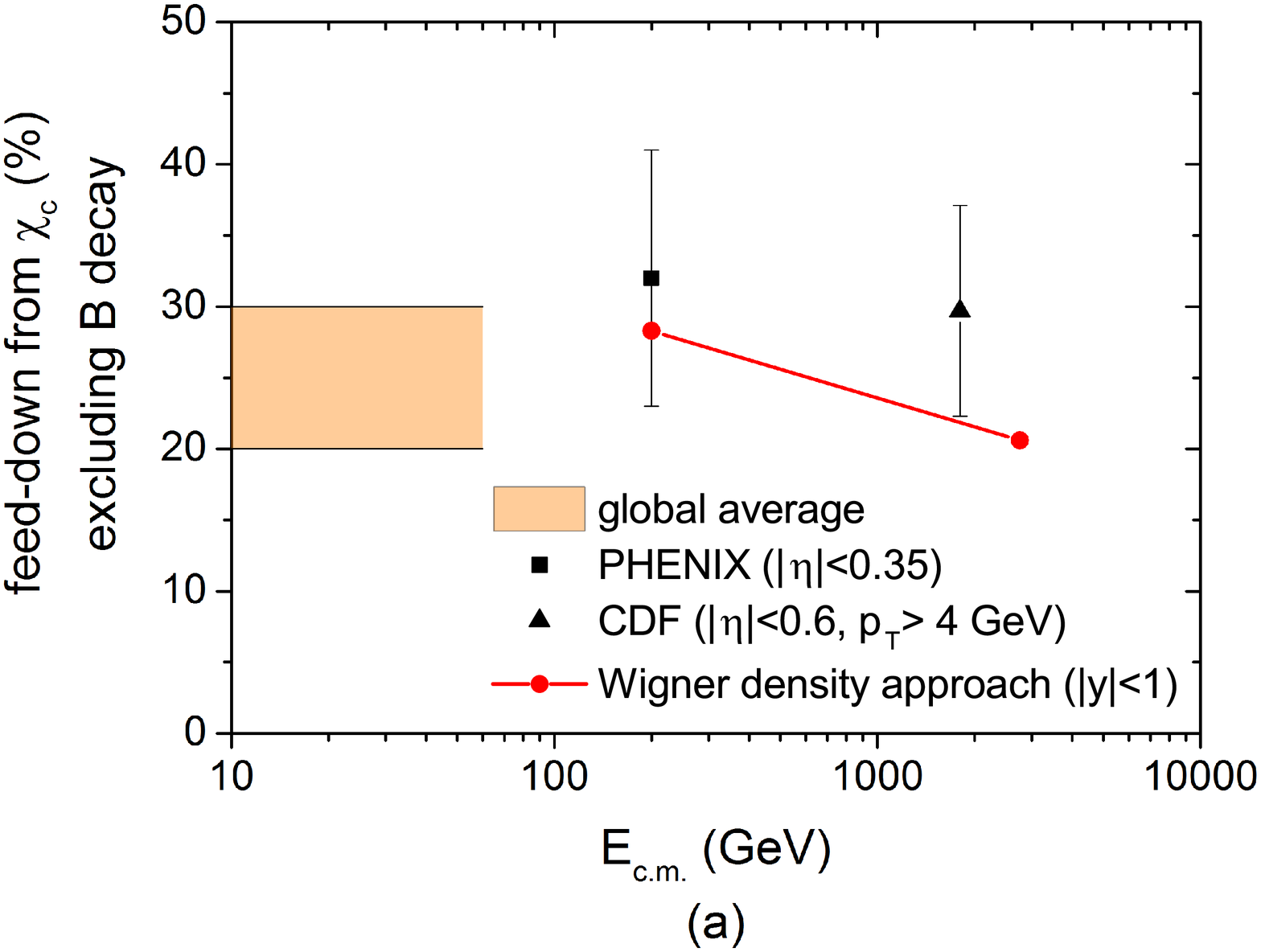}}
\centerline{
\includegraphics[width=9.5 cm]{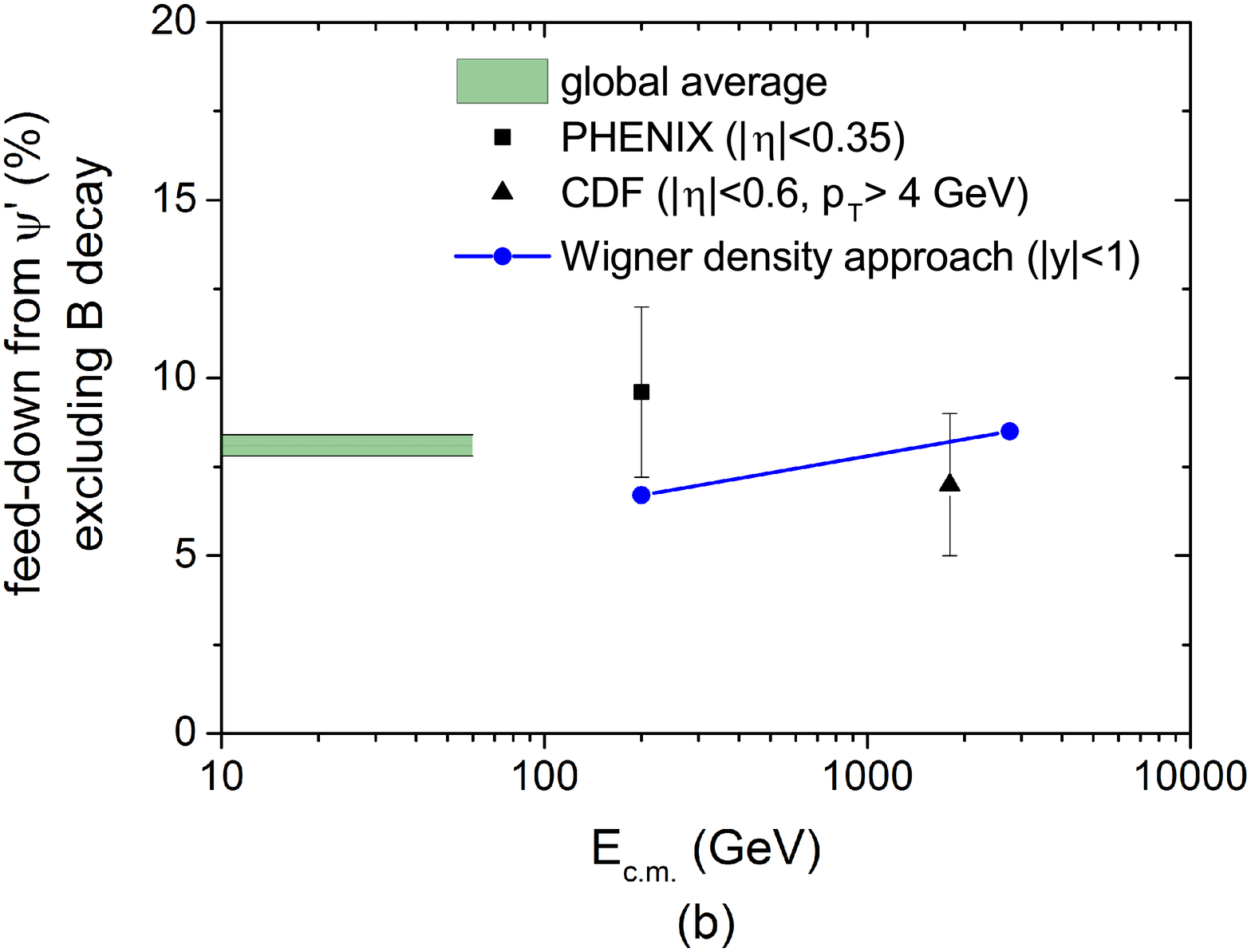}}
\caption{(Color online) The percentages of $J/\psi$ coming from $\chi_c$ and $\psi^\prime$ decays in p+p collisions at $\sqrt{s}=$ 200 GeV and 2.76 TeV, compared with the experimental data from the PHENIX~\cite{Adare:2011vq} and CDF~\cite{Abe:1997yz} collaborations and the global averages of the p+A measurements~\cite{Faccioli:2008ir}. The contribution from $B$ decay is excluded.}
\label{feeddown}
\end{figure}

In Fig.~\ref{feeddown} we compare the percentages of $J/\psi$ coming from $\chi_c$ and $\psi^\prime$ decays in p+p collisions at $\sqrt{s}=$ 200 GeV and 2.76 TeV  with the experimental data from the PHENIX~\cite{Adare:2011vq} and the CDF~\cite{Abe:1997yz} collaboration and with the global averages of the p+A measurements.
The contribution from $B$ decay is excluded as in the experimental data.
We can see that both percentages are reproduced within the error bars of experimental data.
The radii of $J/\psi~(1S)$, $\chi_c~(1P)$ and $\psi^\prime~(2S)$, which are the only parameters in our model, are taken to be 0.5, 0.55, and 0.9 fm, respectively, in order to reproduces the experimental data on total $J/\psi$ as well as on the feed-down from excited charmonia.
They are not far from the radii which are calculated from the potential model~\cite{Karsch:1987zw}.

\section{primordial $J/\psi$ production in heavy-ion collisions}\label{hic}

Motivated by the success in describing the charmonium production in p+p collisions, we apply in this section the same method to obtain the primordial distribution of charminia in relativistic heavy-ion collisions. Also here the initial charm quark pairs are produced in binary  nucleon-nucleon collisions. They are realized by using the Parton-Hadron-String Dynamics (PHSD) which adopts the Monte-Carlo Glauber Model for the charm pair production~\cite{Song:2015sfa,Song:2015ykw}.
The collision energy of each nucleon-nucleon binary collision is smeared out  due to the Fermi momentum in the rest frame of nuclei, but this affects only little the heavy quark production.
According to the simulations, about 16.5 and 108 pairs of charmed quarks  are produced in $0-20$ \% central Au+Au collisions at $\sqrt{s}=$ 200 GeV and in $0-20$ \% central Pb+Pb collisions at $\sqrt{s}=$ 2.76 TeV, respectively, not including the shadowing effect, which causes a considerable suppression of charm production at the LHC energy.
Considering that the radius of a Au or Pb nucleus is $6-7$ fm and that the longitudinal size of the colliding nuclei is extremely small due to the Lorentz contraction, initial charm quark pairs are compactly distributed in a small volume.
Therefore,  primordial charmonium might be produced from two different charm- anticharm pairs.

\begin{figure}[!h]
\centerline{
\includegraphics[width=9.5 cm]{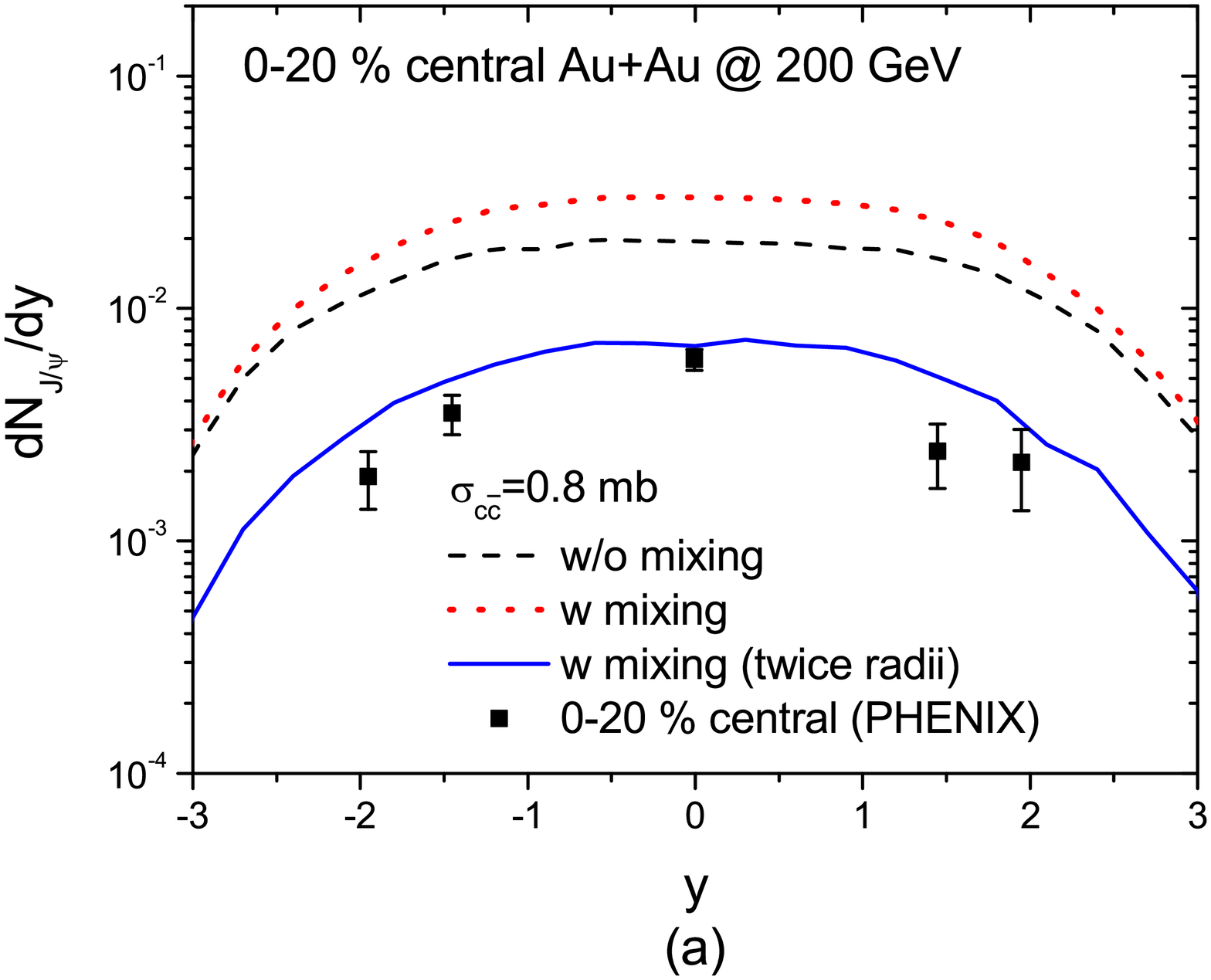}}
\centerline{
\includegraphics[width=9.5 cm]{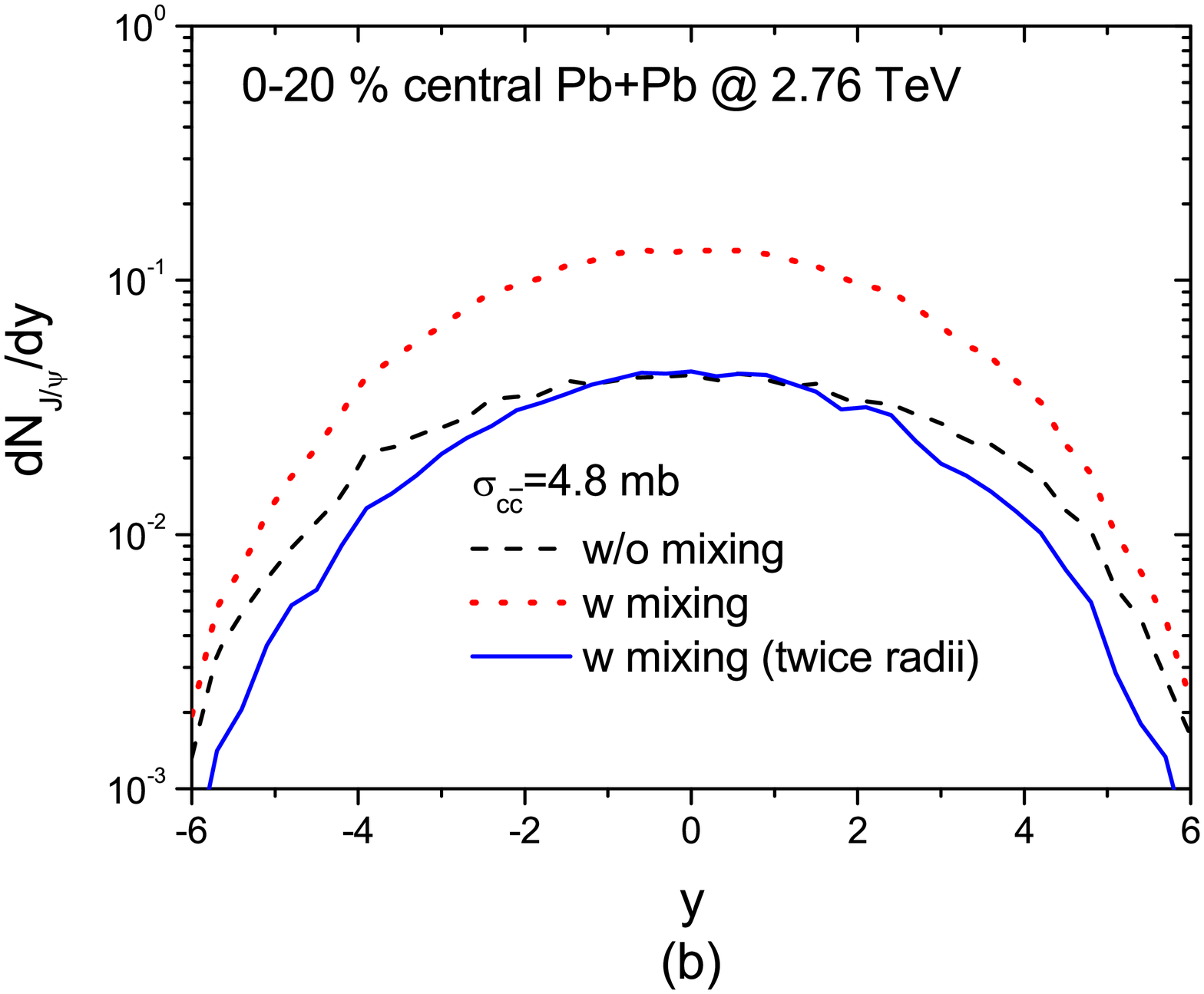}}
\caption{(Color online) Primordial rapidity distributions of $J/\psi$ in $0-20$ \% central Au+Au collisions at $\sqrt{s}=$ 200 GeV (top) and in $0-20$ \% central Pb+Pb collisions at $\sqrt{s}=$ 2.76 TeV (bottom). For the dashed curve the creation of charmonium from
c and $\bar c$ from different initial collisions is excluded.  For the dotted line we do not exclude this mixing. The solid lines show calculations in which the Wigner density has been modified to simulate QGP effects. For orientation we display as well the experimental data from the PHENIX Collaboration~\cite{Adare:2006ns} }
\label{fhic}
\end{figure}

This is shown in figure~\ref{fhic} where the dashed lines are the rapidity distributions  of  the produced $J/\psi$ (including the feed-down from excited states), assuming that charmonium can only be produced from the charm and anti-charm quarks from the same binary nucleon-nucleon collision. This distribution is essentially the same as in Fig.~\ref{rhic} and \ref{lhc} multiplied by the number of binary collisions.

For the dotted lines we allow the formation of charmonium from two different charm-anticharm  pairs.
We find a considerable enhancement of the primordial $J/\psi$  production at both energies.
The production of primordial $J/\psi$ is enhanced by 46 \% in 0-20 \% central Au+Au collisions at $\sqrt{s}=$ 200 GeV, and by a factor of 2.7 in 0-20 \% central Pb+Pb collisions at $\sqrt{s}=$ 2.76 TeV. Since  in the latter collisions more charm pairs are produced, the enhancement of the primordial $J/\psi$ is larger. For the same reason, the enhancement is more prominent in the mid-rapidity region.

When produced in a QGP, according to lattice QCD calculations~\cite{Kaczmarek:2003ph}, the potential energy between charm and anti-charm quarks becomes weaker with increasing temperature of the QGP. Consequently the radii of the charmonia increase or they may become unstable.

In figure~\ref{fhic} we demonstrate the effect of a increasing radius of charmonia on their production probability in relativistic heavy-ion collisions. Larger radii imply larger $\sigma$ values in Eqs.~(\ref{wigner1}) and (\ref{wigner2}). Then only charm quark pairs whose relative momentum is small enough can form a charmonium. On the other hand, it allows that a charm quark pair which is separated by a large distance in coordinate space forms a bound state. Since in heavy-ion collisions charm quarks are already compact in coordinate space, the former effect is stronger than the latter one.
As a result, the production of the primordial $J/\psi$ gets suppressed. Assuming that the radii of charmonia increase by a factor of two, primordial $J/\psi$ decreases to 32 \% in 0-20 \% central Au+Au collisions at $\sqrt{s}=$ 200 GeV, and to 86 \% in 0-20 \% central Pb+Pb collisions at $\sqrt{s}=$ 2.76 TeV, as shown by solid lines in figure~\ref{fhic}.
This implies that QGP effects are important for describing the $J/\psi$ production in relativistic heavy-ion collisions.

That in heavy ion collisions the charmonia are formed in the same way as in p+p collisions is not very realistic, because additional
effects have to be included( and therefore, the experimental data in figure~\ref{fhic} are displayed for orientation only):
Firstly, relativistic heavy-ion collisions produce a QGP which considerably modifies the properties of a charmonium and makes them eventuall unstable. The charm and anti-charm quarks strongly interact with the hot dense QGP matter.
These interactions will change the distribution of charm and anticharm quarks in coordinate and momentum spaces.
This has been theoretically predicted and been proven by many experimental data. They show a nuclear modification factor well below one at high transverse momentum as well as a considerable elliptic flow at low transverse momentum~\cite{Adamczyk:2014uip,Tlusty:2012ix,Adam:2015sza,Abelev:2014ipa,Adare:2006nq,Adare:2014rly}.

Secondly, the excited states and even the ground state of charmonium at LHC energies are most likely dissolved either by color-screening or through thermal decay as long as the dissociation temperature is smaller than the temperature of the plasma.
Therefore the production of the charmonium starts when the expanding plasma has cooled down to the dissociation temperature whose value is presently not known. These charmonia may contain a c and a $\bar c$ comming from the same initial collision but they may be formed as well by a c and a $\bar c$ coming from two different initial pairs.

Additionally, also cold nuclear matter effects change charm production in relativistic heavy-ion collisions. Shadowing suppresses charm production at low transverse momentum and near mid-rapidity. This will directly affect charmonium production. Finally, also hadronic rescattering which breaks up a charmonium into two open charm mesons may suppress the observed yield of quarkonia.

The realistic description of the  time-evolution of charm and anti-charm quarks in matter and of the formation of charmonia in relativistic heavy-ion collisions requires therefore a microscopic approach. This will be studied in near future within the PHSD approach.

\section{summary}\label{summary}

Quarkonium production is special in the respect that it includes both, perturbative and non-perturbative, processes, where the former corresponds to the production of a heavy quark pair and the latter to the formation of a bound state from the pair.
Though the former process is well described by pQCD, the latter needs a model.

In this study, we have shown that charmoninum production in p+p collisions can be described by projecting the position and momentum of the charm and anticharm quarks onto the Wigner density of the charmonium. The energy and momenta of the initial charm and anticharm quarks are given by the PYTHIA event generator, which has been tuned to reproduce the FONLL calculations. The probability for the formation of a $S-$wave charmonium is large if the charm and anticharm quarks are close to each other in coordinate as well as in momentum space. In the case of $P-$wave coalescence, however, the probability peaks at a certain distance. If the probability for $P-$wave charmonium is negative, which happens for a charm quark pair close to each other in coordinate and momentum spaces simultaneously, the probability of a charmonium production is set to zero. The radius of each charmonium state is a parameter to fit the experimental data on the total yield of $J/\psi$ as well as the feed-down from the excited states of charmonium.
We have found that the experimental data in p+p collisions at $\sqrt{s}=$ 200 GeV and 2.76 TeV are reproduced with the rms radii of 0.5, 0.55, and 0.9 fm respectively for $J/\psi~(1S)$, $\chi_c~(1P)$ and $\psi^\prime~(2S)$, which are close to those calculated in a potential model. The distinguished feature of our approach is that it takes into account the spatial information of charm and anti-charm quarks.

While in p+p collisions mostly only one pair of charm quarks is produced, in relativistic heavy-ion collisions many charm quark pairs are produced in a small volume. Since many charm quark pairs are produced nearby in coordinate space, there is a probability for charm and anticharm quarks from two different initial pairs to form a charmonium.
Using the same approach as in p+p collisions but allowing for the formation of charmonium from two different initial charm quark pairs, we have found 46 \% enhancement of primordial $J/\psi$ in 0-20 \% cental Au+Au collisions at $\sqrt{s}=$ 200 GeV and  an enhancement of a factor of 2.7 in 0-20 \% cental Pb+Pb collisions at $\sqrt{s}=$ 2.76 TeV.

The properties of charmonia change in the QGP matter produced in relativistic heavy-ion collisions. According to lattice QCD calculations the potential energy between the charm and anticharm quark becomes less deep at finite temperatures, and consequently the radius of the charmonium increases. To estimate this effect we simply increased the radius by a factor of two
and found that the production of primordial $J/\psi$ reduces to 32 \% in 0-20 \% central Au+Au collisions at $\sqrt{s}=$ 200 GeV, and to 86 \%  in 0-20 \% central Pb+Pb collisions at $\sqrt{s}=$ 2.76 TeV, though the formation of charmonium
by quarks from two different initial collisions is allowed.
It implies that the nuclear matter effects are important to describe $J/\psi$ production in relativistic heavy-ion collisions.

Besides the change of charmonium radii, there are many other effects in relativistic heavy-ion collisions, for example, the strong interaction of charm and anticharm quarks with the QGP, the dissociation of charmonia in QGP, regeneration, and the cold nuclear matter effects such as shadowing.
They will be consistently taken into account in our future study.

\section*{Acknowledgements}
The authors acknowledge inspiring discussions with W.~Cassing,
M. Gorenstein and P. Moreau.
This work was supported by the LOEWE center ``HIC for FAIR" and the convention de financement 2015-08473 de la
region Pays de la Loire. The computational resources have been
provided by the LOEWE-CSC.

\end{document}